\documentclass[fleqn,usenatbib]{mnras}

\usepackage[T1]{fontenc}
\usepackage{ae,aecompl}
\usepackage{color}

\usepackage{graphicx}	
\usepackage{amsmath}	
\usepackage{amssymb}	
\usepackage{gensymb}

\title[Strongly Lensed Quasar SDSS\,J0909$+$4449]{SDSS\,J0909$+$4449: A Large-separation Strongly Lensed Quasar at $z \sim 2.8$ with Three Images}

\author[Shu et al. 2018]{
Yiping Shu,$^{1, 2}$\thanks{E-mail: yiping.shu@ast.cam.ac.uk}\thanks{Royal Society -- K. C. Wong International Fellow}
Rui Marques-Chaves,$^{3, 4}$
N. Wyn Evans,$^{2}$
and Ismael P\'{e}rez-Fournon$^{3,4}$
\\
$^{1}$Purple Mountain Observatory, Chinese Academy of Sciences, 2 West Beijing Road, Nanjing 210008, China\\
$^{2}$Institute of Astronomy, University of Cambridge, Madingley Road, Cambridge CB3 0HA, UK\\
$^{3}$Instituto de Astrof\'{i}sica de Canarias, C/ V\'{i}a L\'{a}ctea, s/n, 38205 San Crist\'{o}bal de La Laguna, Tenerife, Spain\\
$^{4}$Universidad de La Laguna, Dpto. Astrof\'{i}sica, E-38206 La Laguna, Tenerife, Spain\\
}

\date{Accepted XXX. Received YYY; in original form ZZZ}

\pubyear{2018}

\begin{document}
\label{firstpage}
\pagerange{\pageref{firstpage}--\pageref{lastpage}}
\maketitle

\begin{abstract}
We report the discovery of SDSS\,J0909$+$4449, an exceptional system consisting of a quasar at $z=2.788$ strongly lensed by a group of galaxies at $z \sim 0.9$ into three images separated by up to 14\arcsec\, based on archival data collected by the Sloan Digital Sky Survey, extended Baryon Oscillation Spectroscopic Survey, Beijing--Arizona Sky Survey, the Mayall z-band Legacy Survey, and the Gemini Telescope. We discuss two hypotheses on the nature of SDSS\,J0909$+$4449, i.e. a rare triply-imaged quasar in the naked cusp configuration and a typical quadruply-imaged quasar with the fourth image undetected in the current data. We find that simple lens models can provide excellent fits to the observed image positions and the non-detection under either hypothesis. Deeper imaging data, spectroscopic observations, and follow-up light curve measurements will be helpful in determining which hypothesis is correct and provide better constraints on the lens mass distribution. Nevertheless, given its unusually large image separations, SDSS\,J0909$+$4449 will be a unique probe for the mass structure and the underlying cooling and stellar feedback processes on group or cluster scales. 

\end{abstract}




\section{Introduction}

Strong gravitationally lensed quasars with image separations larger than $10^{\prime \prime}$ are valuable 
cosmological probes because they typically require galaxy groups or clusters, which are among the largest 
known structures in the universe, acting as gravitational lenses. A statistical sample of large-separation ($> 10^{\prime \prime}$) strongly lensed quasar systems can therefore be used to probe the abundance and growth history of cosmic 
structures and place constraints on cosmological parameters \citep[e.g.,][]{Narayan88, Turner90, 
Fukugita90, Wambsganss95, Kochanek95b, Kochanek96, Lopes04, Oguri04, Li07, Oguri12b}. 
However, only three of all the known $\sim200$ strongly lensed quasars have maximum image separations larger 
than $10^{\prime \prime}$ \citep[][]{Inada03, Inada06, Dahle13} despite a number of dedicated searches for 
large-separation strongly lensed quasars \citep[e.g.,][]{Maoz97, Ofek01, Zhdanov01, Phillips01, Miller04}. 
Apart from differences in number densities, another important reason is that galaxy groups and clusters, compared to galaxies, are less efficient 
strong lenses because their inner density profiles are relatively flatter \citep[e.g.,][]{Sand08, SLACSVII, 
SLACSX, Sonnenfeld13b, Newman15, Shu15}, while, generally speaking, the average surface mass density of a lens object (with a smooth mass distribution) needs to be higher than a critical density 
\citep[e.g.,][]{Subramanian86, Narayan96} in order to produce multiple lensed images. 

\begin{figure*}
	\includegraphics[width=0.60\textwidth]{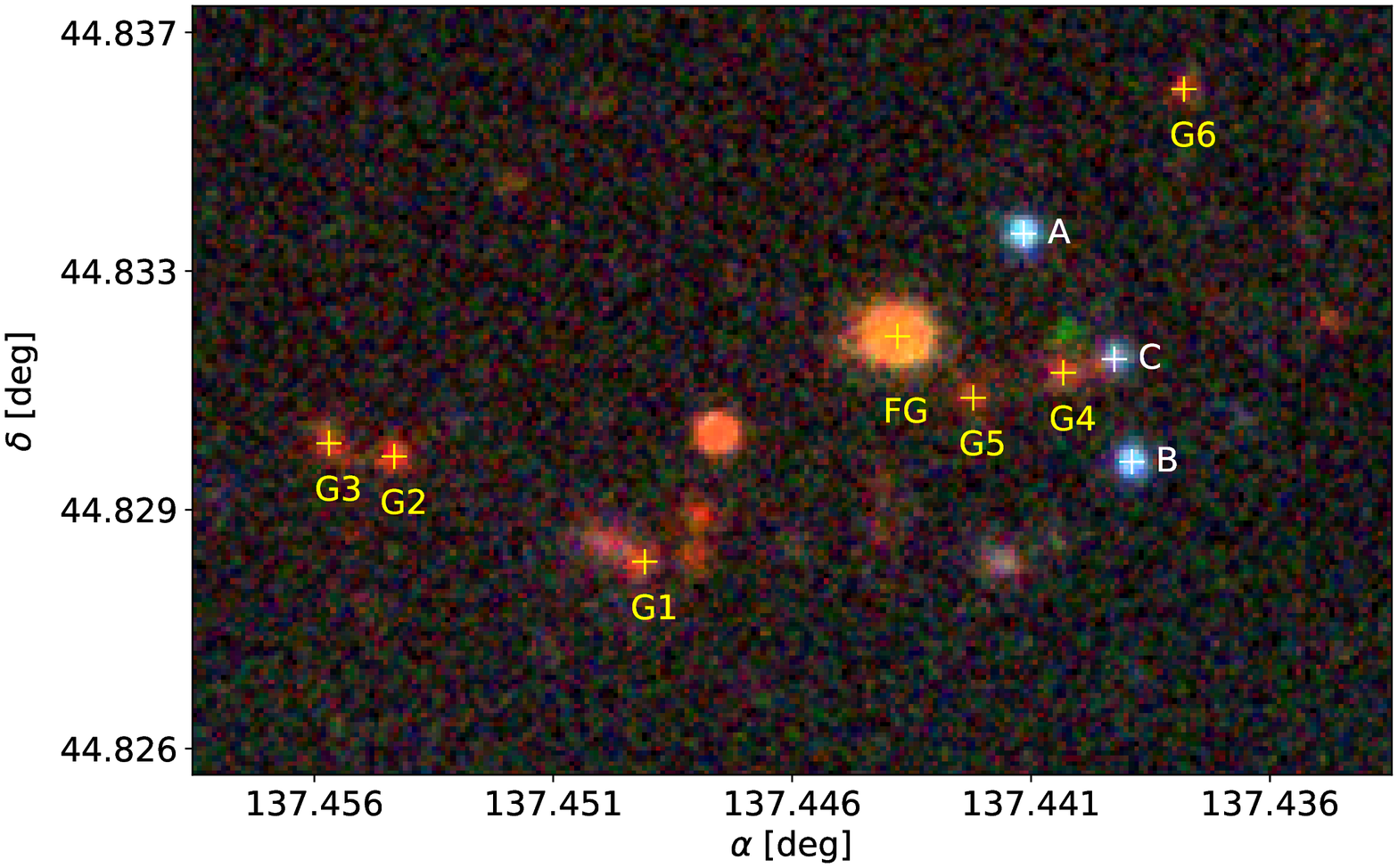}
	\includegraphics[width=0.38\textwidth]{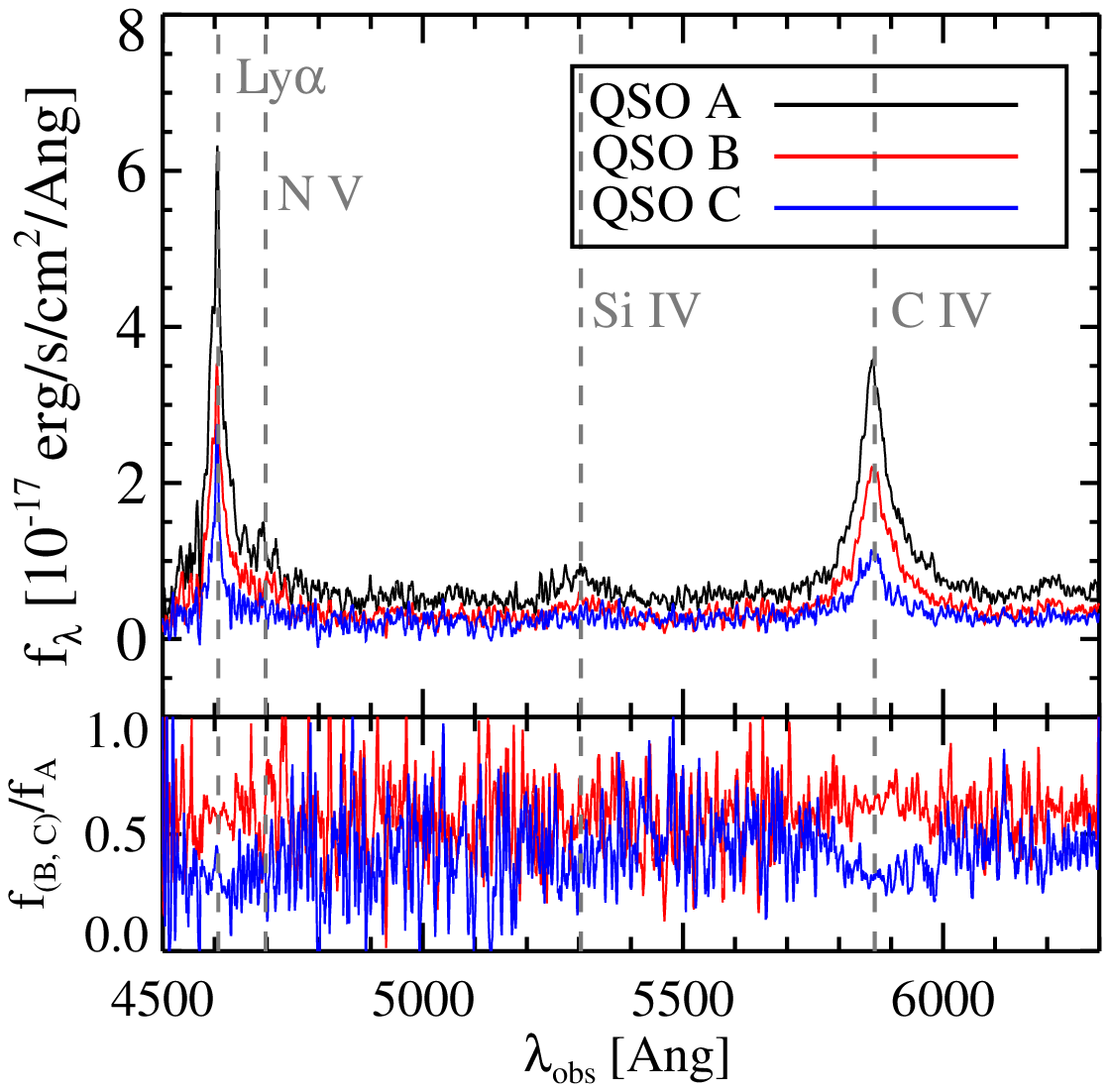}
    \caption{\label{fig:image_spectra} {\it Left}: BASS$+$MzLS color cutout of SDSS\,J0909$+$4449. Three spectroscopically confirmed lensed quasar images are labeled as A, B, and C. Nearby galaxies that are potentially related to the lens are labeled accordingly. {\it Top right}: Gemini GMOS spectra of the three quasar images A (black), B (red), and C (blue). {\it Bottom right}: Ratios of B and C spectra with respect to A. (The green object north of G4 is a known asteroid 2015 TO103.) }
\end{figure*}
\begin{table*}
\centering
\caption{\label{tb:properties} Astrometry and photometry of three quasar images (A, B, and C) and three potential primary lens galaxies (G4, G5, and FG). Positions with respect to the position of C at $\alpha$=09:09:45.34, $\delta$=$+$44:49:53.90 (J2000) are from BASS$+$MzLS data. Redshifts for A, B, C, and FG are spectroscopic, and redshifts for G4 and G5 are photometric. }
\begin{tabular}{c c c c c c c}
\hline
\hline
Object & A & B & C & G4 & G5 & FG \\
\hline
$\Delta \alpha \cos (\delta_C)$ ($^{\prime \prime}$) & $4.97\pm0.03$ & $-0.97\pm0.03$ & $0.00 \pm 0.03$ & $2.80\pm0.05$ & $7.73\pm0.06$ & $11.89\pm0.03$ \\
$\Delta \delta$ ($^{\prime \prime}$) & $6.88\pm0.03$ & $-5.63\pm0.03$ & $0.00 \pm 0.03$  & $-0.75\pm0.05$ & $-2.13\pm0.06$ & $1.25\pm0.03$ \\
Redshift & $2.7887 \pm 0.0004$ & $2.7886 \pm 0.0002$ & $2.788 \pm 0.005$ & $0.76_{-0.04}^{+0.33}$ & $1.03_{-0.16}^{+0.13}$ & $0.43285 \pm 0.00009$ \\
$g_{\rm BASS+MzLS}$ (mag) & $21.59 \pm 0.02$ & $21.66 \pm 0.02$ & $21.78 \pm 0.04$ & $24.19 \pm 0.38$ & $26.86 \pm 3.30$ & $21.13 \pm 0.03$ \\
$r_{\rm BASS+MzLS}$ (mag) & $21.12 \pm 0.03$ & $21.37 \pm 0.03$ & $21.38 \pm 0.05$ & $22.72 \pm 0.18$ & $23.35 \pm 0.24$ & $19.26 \pm 0.01$ \\
$z_{\rm BASS+MzLS}$ (mag) & $21.16 \pm 0.04$ & $21.36 \pm 0.04$ & $20.76 \pm 0.04$ & $20.51 \pm 0.04$ & $21.25 \pm 0.05$ & $18.18 \pm 0.01$ \\
$u_{\rm SDSS}$ (mag) & $22.20 \pm 0.22$ & $21.85 \pm 0.16$ & $23.13 \pm 0.52$ & ... & ... & ... \\
$g_{\rm SDSS}$ (mag) & $21.55 \pm 0.06$ & $20.98 \pm 0.04$ & $22.26 \pm 0.11$ & ... & ... & ... \\
$r_{\rm SDSS}$ (mag) & $21.44 \pm 0.07$ & $20.83 \pm 0.04$ & $22.02 \pm 0.13$ & ... & ... & ... \\
$i_{\rm SDSS}$ (mag) & $21.40 \pm 0.09$ & $20.97 \pm 0.06$ & $21.82 \pm 0.15$ & ... & ... & ... \\
$z_{\rm SDSS}$ (mag) & $21.38 \pm 0.30$ & $21.03 \pm 0.23$ & $22.13 \pm 0.56$ & ... & ... & ... \\
\hline
\hline
\end{tabular}
\end{table*}

Additionally, the image multiplicity in a strong-lens system serves as a useful first-order measure of 
the total mass (dark matter and baryonic matter) distribution of the lens. An interesting and special phenomenon for a lens with 
surface mass density moderately above the critical density, referred to as a ``marginal lens'', is its unique capability 
of producing exactly three lensed images in the so-called naked cusp configuration \citep[e.g.,][]{Kassiola93, Evans98}. 
Theoretical studies have suggested that naked cusp configurations should be more common in group- and cluster-scale lenses than in galaxy-scale lenses, and the fraction of naked cusp configurations in large-separation lenses depends strongly on the central density slope \citep[e.g.,][]{Oguri04b, Minor08}. Considering that dark matter distribution (in a dark-matter only universe) has been well characterized by numerical simulations \citep[e.g.,][]{Navarro97}, 
the image multiplicities can therefore be used to constrain baryonic processes and the interplay between dark matter and baryonic matter that further shape the total mass-density distributions on various mass scales. Similar to large-separation strongly lensed quasars, triply-imaged quasars are also rare with only two such 
discoveries so far \citep{Lewis02, Oguri08}. 

In this Letter, we report the discovery of SDSS\,J0909$+$4449, a strongly lensed quasar at $z \sim 2.8$ with three detected lensed images widely separated by up to $\sim 14^{\prime \prime}$. 
Section~\ref{sect:data} presents photometric and spectroscopic properties of SDSS\,J0909$+$4449. Section~\ref{sect:interp} provides lens models under two possible hypotheses, i.e. a triply-imaged lens or a quadruply-imaged lens with the fourth image too faint to be detected in the current data. Discussions and conclusions are given in Sections~\ref{sect:discussions} and \ref{sect:conclusions}. 

\begin{figure*}
\centering
	\includegraphics[width=0.48\textwidth]{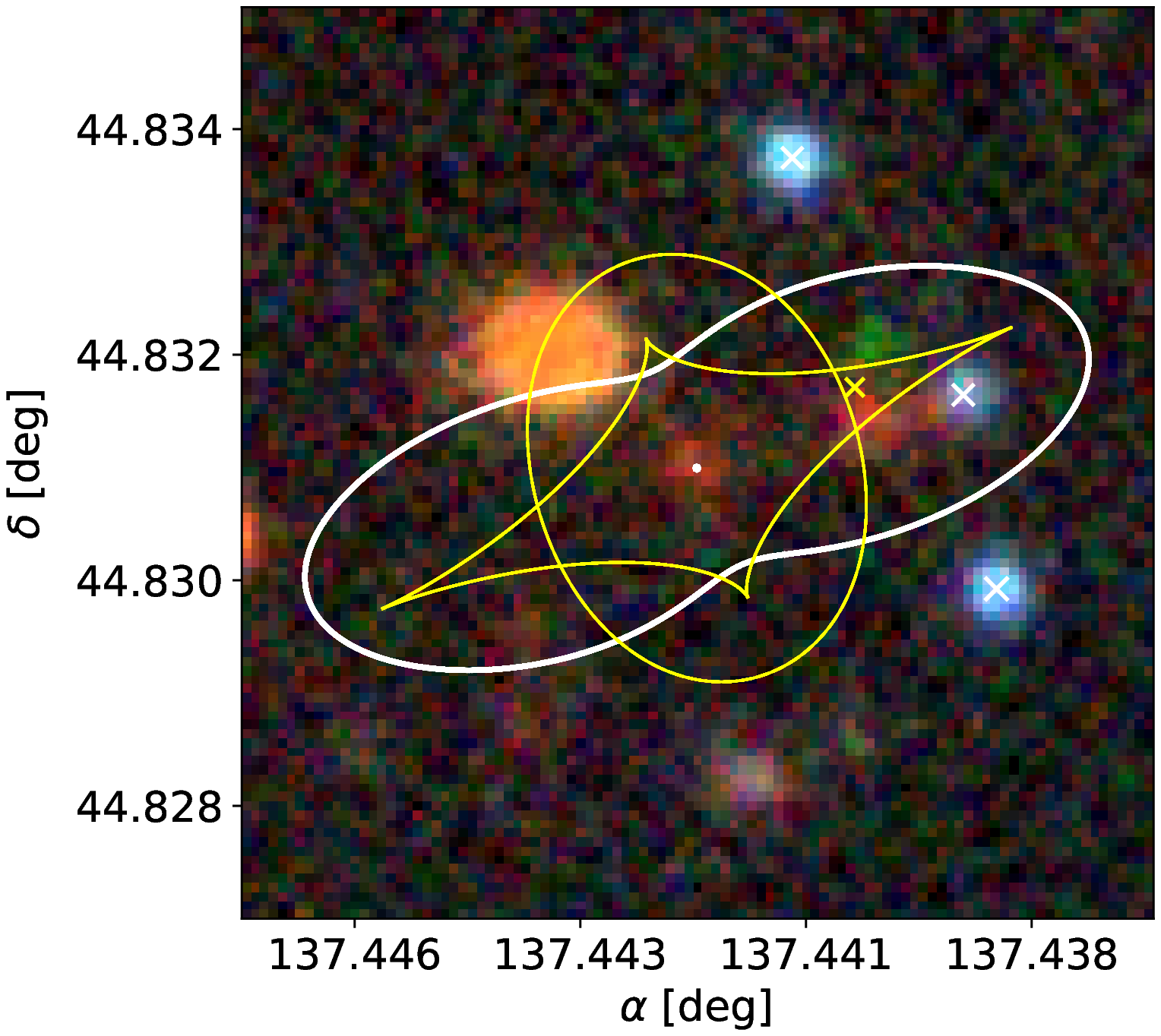}
	\includegraphics[width=0.48\textwidth]{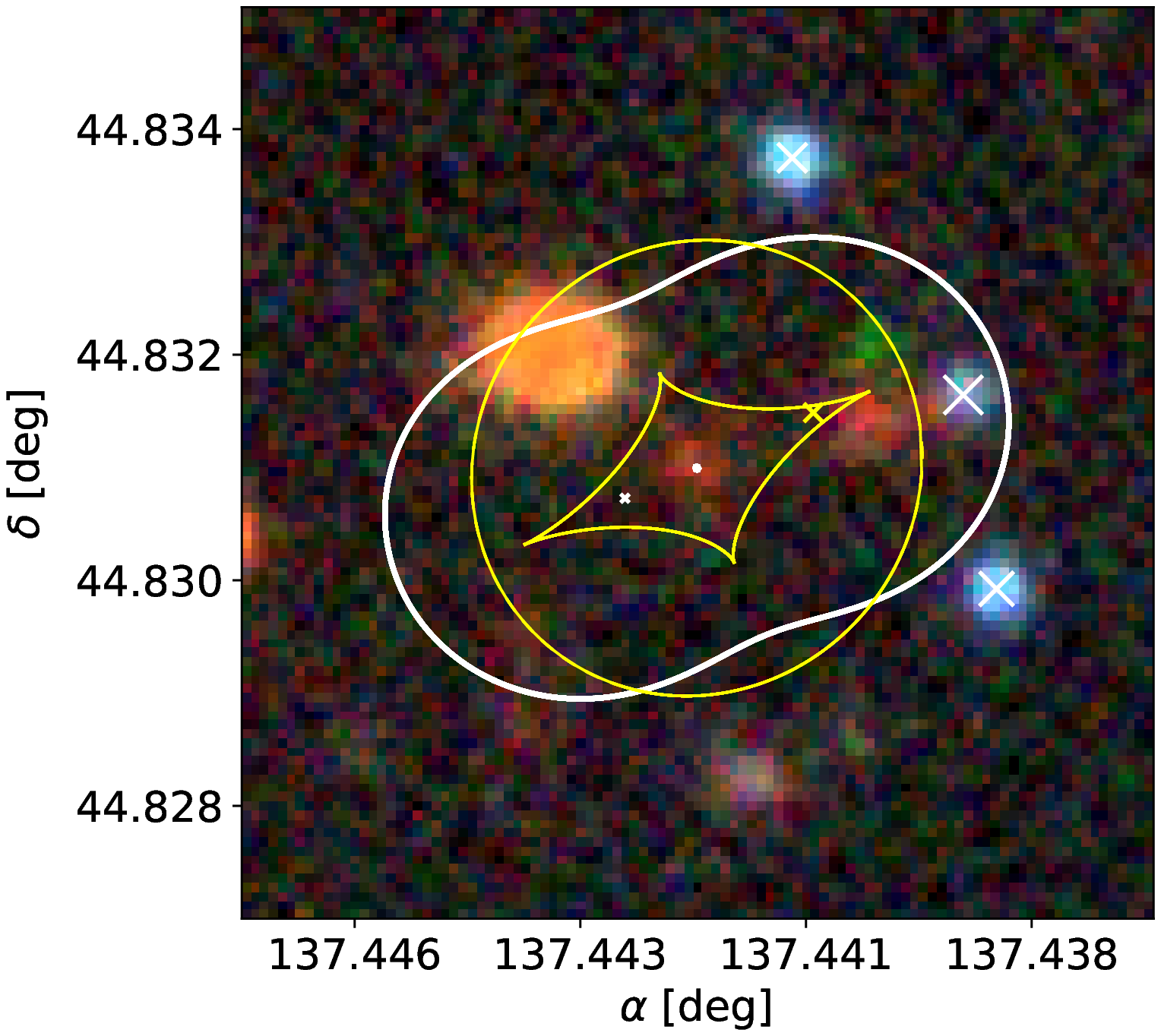}
	\caption{\label{fig:model} Fitting results of the 3-image (left) and 4-image (right) lens models on SDSS\,J0909$+$4449. White crosses mark predicted positions of multiple lensed images with sizes proportional to the lensing magnifications. Yellow crosses indicate the positions of the quasar in the source plane. White and yellow solid lines correspond to critical lines and caustics. Both lens models provide excellent fits to the observed quasar positions.}
\end{figure*}

\section{Data}
\label{sect:data}

SDSS\,J0909$+$4449 was first noticed as a potential strongly lensed quasar system because it contains a quasar pair with same redshifts and nearly identical spectral shapes as measured by the extended Baryon Oscillation Spectroscopic Survey \citep[eBOSS,][]{Dawson16}. 
Specifically, SDSS\,J090945.81$+$445000.6 (denoted as quasar A hereafter) and SDSS\,J090945.25$+$444948.1 (denoted as quasar B hereafter) were spectroscopically confirmed as two quasars at $z_{\rm A}=2.7887 \pm 0.0004$ and $z_{\rm B}=2.7886 \pm 0.0002$ respectively with a projected separation of 13\farcs86 (i.e. 109 kpc). Imaging data from Sloan Digital Sky Survey \citep[SDSS,][]{York00}, the Beijing-Arizona Sky Survey \citep[BASS,][]{Zou17, Zou17b} and the Mayall z-band Legacy Survey \citep[MzLS,][]{Dey18} further indicate the presence of a third object SDSS\,J090945.33$+$444954.0 (denoted as C hereafter) between A and B with a similar blue color (Figure~\ref{fig:image_spectra}), which was not observed spectroscopically by the eBOSS survey. A, B, and C are aligned to form an arc-like configuration. A search in the data archive found that A, B, and C were all observed spectroscopically in 2012 by the Gemini Multi-Object Spectrograph (GMOS) on the Gemini Telescope (PI: Adam Myers). The reduced GMOS spectra in Figure~\ref{fig:image_spectra} clearly show that object C is also a quasar with a spectral shape very similar to that of A or B. A simple fit to the GMOS spectrum of C gives $z_{\rm C}=2.788 \pm 0.005$.  
The imaging and spectroscopic data jointly suggest that quasars A, B, and C are lensed images of a single quasar at $z=2.788$. Their relatively large separations imply a group-scale or cluster-scale lens. This is supported by the presence of several extended objects with similar red colors that are classified as galaxies morphologically in the Data Release 6 (DR6) of the Legacy Surveys, namely G1--G6 (ordered by their $z$-band brightness). Analyses by EAZY \citep[a photometric redshift code by][]{Brammer08} suggest that the photometric redshifts of G1--G6 are from 0.7 to 1.0 with a typical uncertainty of 0.1. Galaxy FG is spectroscopically confirmed as an elliptical galaxy in the foreground ($z=0.433$) by the BOSS survey \citep{Dawson13}. 

Table~\ref{tb:properties} presents basic properties of A, B, C, and nearby galaxies that are potentially related to the lens. Here and henceforth, all magnitudes are given in the AB system. Positions (with respect to C) are taken from the Tractor Catalog in DR6 of the Legacy Surveys. The redshift row provides spectroscopic redshifts for A, B, C, and FG, and photometric redshifts for G4 and G5. BASS$+$MzLS $grz$ magnitudes and SDSS $ugriz$ data are provided. We choose not to report SDSS magnitudes for nearby galaxies that are generally too faint to be detected or accurately measured by SDSS. The quasar intrinsic variabilities clearly manifest in these two datasets. Magnitudes of A stayed roughly unchanged between the two epochs. However, B was the brightest among the three lensed images in the SDSS epoch (in 2001), while A became the brightest in the BASS$+$MzLS epoch (in 2016--2017). C was the faintest image in both epochs. 
Some red features are noticed in the BASS$+$MzLS data to be close to and presumably overlap with C, which could 
contaminate the $z$-band magnitude of C. The average $g-r$ and $r-z$ colors of A, B, and C are 
$0.38 \pm 0.05$ and $-0.02 \pm 0.04$. The typical $g-r$ and $r-z$ colors of the nearby galaxies are 
1.4 and 2.2 respectively. 

\section{Interpretations}
\label{sect:interp}

Interestingly, three, instead of the more commonly seen two or four, lensed images are detected for SDSS\,J0909$+$4449. A conservative search in the Tractor Catalog for ``blue'' objects with $g-r<0.6$ and $r-z < 0.7$ (color cuts satisfied by A, B, and C) within 60$^{\prime \prime}$ of C only finds an extended galaxy and three faint objects west of C, none of which can be naturally interpreted as a lensed quasar image. In this section, we present two possible hypotheses regarding the nature of SDSS\,J0909$+$4449. Considering the significant intrinsic variabilities of the quasar images, we choose to only use positions of the three quasar images in our modelling, so in total we have 6 constraints. 

\subsection{Hypothesis One -- A Triply-Imaged Lens}

We first consider SDSS\,J0909$+$4449 as a strongly lensed quasar with exactly three lensed images in the naked cusp configuration, which would make it the third naked-cusp quasar lens after APM 08279$+$5255 \citep{Lewis02} and SDSS\,J1029$+$2623 \citep{Oguri08}. 

To examine this hypothesis, we consider a simple lens model consisting of an singular isothermal ellipsoid (SIE) mass distribution in an external shear field with 7 free parameters (assuming the SIE mass component and the external shear field is co-centered). The quasar position in the source plane are two other free parameters. In total, this lens model has 9 free parameters. To mitigate the problem of having more free parameters than constraints, we consider three sets of lens models with centers of the mass component fixed to the light centers of G4, G5, and FG, three galaxies that could possibly be the primary lens object, and use the {\tt lensmodel} toolkit \citep{Keeton01} to verify whether any set of the model can fit the data. The principal conclusions of the modelling have been reproduced with {\tt Pixelens} \citep{Saha11} as an additional check.

We find that only the lens model with mass center fixed at G5 can produce exactly three lensed images at locations that match the data well ($\chi^2 \sim 10^{-5}$). The left panel in Figure~\ref{fig:model} shows the results of this 3-image model overlaid on the observed imaging data. Three white crosses show the model-predicted positions of the three lensed images with the size of the cross proportional to the lensing magnification. The yellow cross indicates the position of the quasar in the source plane. The white and yellow lines correspond to the critical line and caustics. This model suggests that the Einstein radius is 5\farcs0, the axis ratio of the SIE component is 0.43, and the strength of the external shear field is 0.26. The position angle of the external shear field (120$^{\circ}$ east of north) is consistent with the spatial distribution of nearby galaxies. A is predicted to be the leading image followed B and C, respectively. The model-predicted magnifications are $\mu_{\rm A}=2.2$, $\mu_{\rm B}=2.6$, and $\mu_{\rm C}=2.3$, respectively.  

\subsection{Hypothesis Two -- A Quadruply-Imaged Lens}

Another possibility is that SDSS\,J0909$+$4449 is a typical quadruply-imaged system with the fourth image, denoted as D, too faint to be detected in the BASS$+$MzLS data. Considering that the limiting $g$-band magnitude of BASS$+$MzLS data is 24.7 \citep{Schlegel15}, it would require D to be fainter than A by a factor of $\sim$16. 

\begin{figure}
\centering
	\includegraphics[width=0.45\textwidth]{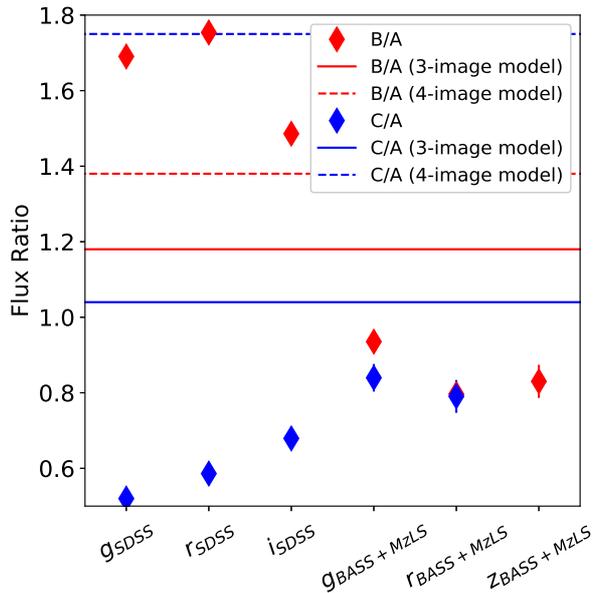}
	\caption{\label{fig:fluxratios} Observed flux ratios B/A (red diamonds) and C/A (blue diamonds) for SDSS\,J0909$+$4449 in various filter bands in the SDSS epoch and BASS$+$MzLS epoch. Error bars for most of the data points are smaller than the size of the diamond. Solid red (blue) line indicates the predicted flux ratio B/A (C/A) by the 3-image model, and dashed lines show the same predictions by the 4-image model.}
\end{figure}

We again use the SIE$+$external shear lens model with mass center fixed at G5 to test this hypothesis. We assume D is located at (30\arcsec, 0\arcsec) east of C with a positional uncertainty of 30\arcsec\, in each direction, and optimize the lens model with these two extra ``constraints''. We obtain a good fit with $\chi^2=0.5$ for 1 degree of freedom. The right panel in Figure~\ref{fig:model} shows results of this 4-image model. The small white cross $\sim$2\arcsec\, east of G5 indicates the predicted position of D. This model suggests that the Einstein radius is 6\farcs6, the axis ratio of the SIE component is 0.86, the strength of the external shear is 0.32, and the position angle of the external shear is 117$^{\circ}$ (east of north). A is the leading image followed by B, C, and D, respectively. Different from the 3-image model above, the 4-image model predicts that C, instead of B, should be the most magnified image. The model-predicted magnifications are $\mu_{\rm A}=4.1$, $\mu_{\rm B}=5.6$, $\mu_{\rm C}=7.1$, and $\mu_{\rm D}=0.4$, respectively. 
The A/D ratio is roughly consistent with the non-detection after taking into account intrinsic variabilities of the quasar (Figure~\ref{fig:fluxratios}, a factor of $\sim 2$).

A lens model with mass center fixed at G4 can also provide a good fit to the data, but with a slightly larger $\chi^2$ value (0.8). Moreover, it requires an unusually strong external shear (strength=0.57) and predicts D to be $40\%$ brighter than in the above model, which make this model less favorable. A lens model with mass center fixed at FG fails to provide a good fit. 

\section{Discussion}
\label{sect:discussions}

First of all, we emphasize that SDSS\,J0909$+$4449 is confirmed as a strongly lensed quasar. Although only simple lens models are built here given the limitations of the current data, they convincingly establish the lensing hypothesis.

Blue (red) diamonds in Figure~\ref{fig:fluxratios} show the observed B/A (C/A) in SDSS $gri$ bands and BASS$+$MzLS $grz$ bands, and the blue (red) solid and dashed lines indicate the predicted B/A (C/A) under two hypotheses. The flux ratios in SDSS $u$ and $z$ bands are not shown due to relatively large photometric uncertainties, and the C/A flux ratio in the BASS$+$MzLS $z$ band is not shown because the $z$-band magnitude of C is significantly affected by nearby, red features (likely a galaxy in the foreground). The flux ratios do not show significant variations in different filter bands within the same epoch. The average B/A flux ratio was 1.64 with a scatter of 0.11 in the SDSS epoch, and became 0.85 with a scatter of 0.06 in the BASS$+$MzLS epoch. Considering that A had similar magnitudes across the two epochs, it indicates that B had varied in brightness by a factor of $\sim2$ in this period. The average C/A flux ratio was 0.60 with a scatter of 0.06 in the SDSS epoch, and became 0.82 with a scatter of 0.02 in the BASS$+$MzLS epoch. The differences between the model-predicted and observed flux ratios in SDSS\,J0909$+$4449 under the two hypotheses are both consistent with values found in strongly lensed quasars in general \citep[e.g.,][and references therein]{Pooley07}. In addition, the magnification of C is presumably affected by the nearby, foreground galaxies that have not been taken into account in our simple lens models, so one can not yet draw a definitive conclusion on whether SDSS\,J0909$+$4449 is triply or quadruply imaged given current data.

Follow-up observations will help to determine which hypothesis is correct and provide more insight into this unique lens system. We assume that the quasar can vary by up to 1.3 mag, which is five times the largest root-mean-square magnitude change of quasars found in \citep{MacLeod12}. Further considering that the 4-image model predicts D/A=0.09, we suggest that imaging data $\sim$1.5 mag deeper than the BASS$+$MzLS data could confirm whether the fourth quasar image exists. Spectroscopic observations of the foreground galaxies, i.e. G1--G6, will unambiguously confirm if these galaxies indeed form a galaxy group. Accurate spectroscopic redshifts of the lens galaxies in SDSS\,J0909$+$4449 can also lead to accurate estimations on the total mass and dark-matter fraction of the lens.  In addition, deep imaging data may even reveal lensed images of the quasar host galaxy and strongly or weakly lensed images of other background sources, which will provide better constraints on the lens model and hence the mass distribution of the lens. 

\section{Conclusions}
\label{sect:conclusions}

In this Letter, we present an analysis of SDSS\,J0909$+$4449, a system consisting of three $z=2.788$ quasars with nearly identical spectral shapes that are separated by up to 14\arcsec\, in projection. We confirm that SDSS\,J0909$+$4449 is a strongly lensed quasar because the positions of the three quasar images can be well reproduced by simple lens models. A group of galaxies at redshifts of 0.7--1.0 are believed to be associated with the lens. Among $\sim$200 known strongly lensed quasars to date, SDSS\,J0909$+$4449 is the fourth system to have a maximum image separation larger than 10\arcsec. Based on current data, it is still unclear whether SDSS\,J0909$+$4449 is triply imaged in a naked-cusp configuration, which makes it the third strongly lensed quasar with exactly three lensed images, or quadruply imaged with the fourth lensed image undetected. Follow-up imaging and spectroscopic observations will reveal more detailed aspects of SDSS\,J0909$+$4449 including its total mass, dark-matter fraction, and mass distribution in the central region, which will provide insight into the growth history and interplay between dark matter and baryonic matter in massive cosmic structures. 

\section*{Acknowledgements}

We thank the referee for a prompt and constructive report, and Matt Auger, Vasily Belokurov, Sergey Koposov, Xianzhong Zheng, and Cameron Lemon for helpful discussions. Y.S. has been supported by the National Natural Science Foundation of China (No. 11603032 and 11333008), the 973 program (No. 2015CB857003), and the Royal Society -- K.C. Wong International Fellowship (NF170995). RMC and IPF have been supported by the Spanish research grants ESP2015-65597-C4-4-R and ESP2017-86852-C4-2-R.




\bibliographystyle{mnras}
\bibliography{references_db} 


\bsp	
\label{lastpage}
\end{document}